\documentclass[pra,showpacs,twocolumn,superscriptaddress]{revtex4}
\usepackage{epsfig}
\usepackage{amsmath}
\usepackage{graphicx}
\usepackage{color}
\usepackage{float}
\usepackage{ulem}

\large
\newcommand{\be}{\begin{equation}}
\newcommand{\ee}{\end{equation}}
\newcommand{\ba}{\begin{array}}
\newcommand{\ea}{\end{array}}
\newcommand{\bea}{\begin{eqnarray}}
\newcommand{\eea}{\end{eqnarray}}

\def\ket#1{\left| #1\right\rangle}

\begin{document}

\title{A quantum information theoretic analysis of three flavor neutrino oscillations
}

\author{Subhashish Banerjee}
\email{subhashish@iitj.ac.in}
\affiliation{Indian Institute of Technology Jodhpur, Jodhpur 342011, India}

\author{Ashutosh Kumar Alok}
\email{akalok@iitj.ac.in}
\affiliation{Indian Institute of Technology Jodhpur, Jodhpur 342011, India}

\author{R. Srikanth}
\email{srik@poornaprajna.org}
\affiliation{Poornaprajna Institute of Scientific Research, Sadashivnagar,  Bengaluru   560080,  India}

\author{Beatrix C. Hiesmayr}
\email{beatrix.hiesmayr@univie.ac.at}
\affiliation{University of Vienna, Faculty of Physics, Boltzmanngasse 5, 1090 Vienna, Austria}

\title{A quantum information theoretic analysis of three flavor neutrino oscillations
}

\begin{abstract}
Correlations  exhibited  by  neutrino  oscillations  are  studied  via
quantum information  theoretic quantities. We show  that the strongest
type of  entanglement, \textit{genuine} multipartite  entanglement, is
persistent in the  flavour changing states. We prove  the existence of
Bell type nonlocal features, in both  its absolute and genuine
avatars. Finally, we show that a measure of nonclassicality,
 dissension,  which is a  generalization of quantum discord  to the
tripartite case, is nonzero for almost the entire range of time in the
evolution  of   an  initial   electron-neutrino.  Via   these  quantum
information  theoretic  quantities   capturing  different  aspects  of
quantum correlations, we elucidate the differences between the flavour
types,  shedding light  on the  quantum-information theoretic  aspects of  the weak
force.

\end{abstract}
\maketitle
\section{Introduction}
\label{intro}

The study of correlations in quantum systems has a vast literature and
draws its practical importance  from potential applications to quantum
technologies   such   as   quantum  cryptography   and   teleportation
\cite{nc}. Recently, there  has been a move  towards extending these
studies     to    systems     in     the     domain    of     particle
physics~\cite{meson1,meson2,meson3,meson4,meson5,meson6,meson7,meson8,meson9,meson10,meson11,expt1,expt2,expt3,neutri1,neutri2,neutri3,neutri4,neutri5,q2flavn}. Neutrino is
a particularly  interesting candidate  for such  a study.   In nature,
neutrinos are available  in three flavors, viz,  the electron-neutrino $\nu_e$,
muon-neutrino $\nu_\mu$ and  tau-neutrino $\nu_\tau$. Owing  to their non-zero
mass,  they oscillate  from  one  flavor to  another.   This has  been
confirmed  by  a  plethora  of experiments,  using  both  natural  and
``man-made'' neutrinos.

Neutrino     oscillations    are     fundamentally    three     flavor
oscillations. However, in  some cases, it can be  reduced to effective
two  flavor oscillations~\cite{q2flavn}.  These  elementary particles
interact  only  via  weak  interactions, consequently  the  effect  of
decoherence,  as  compared  to  other particles  widely  utilized  for
quantum information  processing, is  small. Numerous  experiments have
revealed     interesting     details     of     the     physics     of
neutrinos~\cite{Bahcall:2004ut,Araki:2004mb,Ashie:2004mr,Michael:2006rx,
Abe:2013hdq,Abe:2013fuq}. This  paper asks  what type
of  quantum correlations  is persistent  in the  time evolution  of an
initial $\nu_e$ or  $\nu_\mu$ or $\nu_\tau$. It  presents a systematic
study of  the many-faceted aspect of  quantum correlations.  Herewith,
it  contributes  to the  understanding  how  Nature processes  quantum
information in the regime of  elementary particles and, in particular,
which aspect  of quantum information  is relevant in  weak interaction
processes.

Three-flavor neutrino oscillations can be studied by mapping the state
of  the  neutrino, treating it as  a  three-mode  system,  to that  of  a
three-qubit system~\cite{neutri1,neutri2}.  In particular,
it  was  shown that  the  neutrino  oscillations  are related  to  the
multi-mode  entanglement  of  single-particle   states  which  can  be
expressed in  terms of flavor  transition probabilities. Here  we take
the  study  of  such  foundational issues  further  by  characterizing
three-flavor neutrino  oscillations by  quantum correlations.  This is
non-trivial as  quantum correlations  in three-qubit systems  are much
more involved compared to their two-qubit counterparts.

The present  study of  quantum correlations  in three  flavor neutrino
oscillations can be broadly classified into three categories:
\begin{itemize}
\item Entanglement: We study various types of the in-separability properties of the dynamics of neutrino oscillations
via the von Neumann entropy  and in terms of a nonlinear witness  of  \textit{genuine} multipartite  entanglement introduced in Ref.~\cite{HMGH}.
\item  \textit{Genuine} multipartite  nonlocality:  Nonlocality -- that is considered to be the  strongest
  manifestation of quantum  correlations -- is studied in both its
  absolute and genuine tripartite  facets, characterized by the Mermin inequalities~\cite{M90}
  and Svetlichny inequalities~\cite{Svet87}.
\item Dissension:  A tripartite generalization  of quantum
  discord which is a measure of nonclassicality of correlations~\cite{chakrabarty}.
\end{itemize}

The plan of the paper is as follows. In Sec.~\ref{sec:1}, we provide a
brief introduction to the phenomenology of neutrinos and introduce the
three-flavour mode  entangled state  which will  be analysed using
information  theoretic  tools.  The  core  of  the  paper  is  Sec.~\ref{sec:2}, where we characterize  three flavor neutrino oscillations
in terms of  various facets of quantum correlations.  We then conclude
by providing an outlook.

\section{Three flavour neutrino oscillations}
\label{sec:1}

The   three   flavours   of  neutrinos,   $\nu_e$,   $\nu_{\mu}$   and
$\nu_{\tau}$, mix to form three  mass eigenstates $\nu_1$, $\nu_2$ and
$\nu_3$:
\begin{equation}
\left(\begin{array}{c}     \nu_e    \\    \nu_{\mu}
  \\  \nu_{\tau}\end{array}\right)  = U  \left(\begin{array}{c}  \nu_1
  \\  \nu_2 \\\nu_3\end{array}\right)\,,  \end{equation}  where $U$  is the $3\times3$  PMNS (Pontecorvo-Maki-Nakagawa-Sakata)
  mixing matrix parameterized by three mixing angles ($\theta_{12}$,
  $\theta_{23}$ and $\theta_{13}$) and a $CP$ violating phase $\delta$ ($C$\dots charge conjugation, $P$\dots parity).
  Neglecting the $CP$ violating phase (that has not been observed) the mixing matrix can be written as
\begin{equation}
U =
\left(\begin{array}{ccc}
c_{12} c_{13} & s_{12} c_{13} & s_{13}  \\
- s_{12} c_{23}- c_{12} s_{23} s_{13} & c_{12} c_{23}- s_{12} s_{23} s_{13}  & s_{23} c_{13}  \\
 s_{12} s_{23}- c_{12} c_{23} s_{13}& - c_{12} s_{23}- s_{12} c_{23} s_{13}  & c_{23} c_{13}
\end{array}\right)\,,
\label{pmns}
\end{equation}
where   $c_{ij}$ and $s_{ij}$ denote $\cos \theta_{ij}$ and $\sin \theta_{ij}$, respectively.

Therefore, each flavor state is given by a linear superposition of the mass eigenstates,
\begin{equation}
\ket{\nu_{\alpha}} = \sum_k U_{\alpha k} \ket{\nu_k}\,,
\end{equation}
where $\alpha = e, \mu, \tau$;  $k = 1,2,3$.
As the massive neutrino states $\ket{\nu_k}$ are eigenstates of the Hamiltonian with energy eigenvalues $E_k$,
the time evolution of the mass eigenstates $\ket{\nu_{k}}$ is given by
\begin{equation}
\ket{\nu_{k}(t)} = e^{-\frac{i}{\hbar} E_{k} t} \ket{\nu_{k}}\,,
\end{equation}
where $\ket{\nu_{k}}$ are the mass eigenstates at time $t=0$.

Straightforwardly, the time evolution of flavor neutrino states computes to
 \begin{equation} \ket{\nu_{\alpha}(t)} = a_{\alpha  e} (t)\ket{\nu_e} +
  a_{\alpha    \mu}     (t)\ket{\nu_{\mu}}    +     a_{\alpha    \tau}
  (t)\ket{\nu_{\tau}}\,,
\label{t2}
\end{equation}
with
\begin{equation}
a_{\alpha \beta} (t) = \sum_k U_{\alpha k}\, e^{-\frac{i}{\hbar} E_k t}\, U^{*}_{\beta k}\,.
\end{equation}

For example, if an electron-neutrino is produced at time
$t=0$  then its  time evolution  is  given by

\begin{equation} \ket{\nu_{e}(t)}  =
a_{ee}   (t)\ket{\nu_e}  +   a_{e\mu}(t)\ket{\nu_{\mu}}  +   a_{e\tau}
(t)\ket{\nu_{\tau}}\,,
\label{e2}
\end{equation}
where
\begin{eqnarray}
a_{ee} (t) &=& |U_{e1}|^2 e^{-\frac{i}{\hbar}  E_1 t} + |U_{e2}|^2 e^{-\frac{i}{\hbar}  E_2 t} + |U_{e3}|^2 e^{-\frac{i}{\hbar}  E_3 t}, \nonumber\\
a_{e\mu} (t) &=& U_{e1} U_{\mu 1}^* e^{-\frac{i}{\hbar}  E_1 t} + U_{e2} U_{\mu 2}^* e^{-\frac{i}{\hbar}  E_2 t} + U_{e3} U_{\mu 3}^* e^{-\frac{i}{\hbar}  E_3 t}, \nonumber\\
a_{e\tau} (t) &=& U_{e1} U_{\tau 1}^* e^{-\frac{i}{\hbar}  E_1 t} + U_{e2} U_{\tau 2}^* e^{-\frac{i}{\hbar}  E_2 t} + U_{e3} U_{\tau 3}^* e^{-\frac{i}{\hbar}  E_3 t}\;.\nonumber
\end{eqnarray}

If we assume that the detected neutrinos have an energy of at least one
MeV (the electron/positron mass), namely are in the ultrarelativistic regime, the flavor eigenstates are well
defined in the context of quantum mechanics~\cite{neutri1}. In this approximation the survival probabilities take the form
\begin{equation}\label{surviveprob}
P_{\nu_{\alpha} \to \nu_{\alpha}} = 1-4 \sum_{k>j} |U_{\alpha k}|^2 |U_{\alpha j}|^2 \sin^2\left(\frac{\Delta m^2_{kj} c^4}{4\hbar c }\frac{L}{E}\right),
\end{equation}
and the oscillation probabilities
\begin{eqnarray}\label{oscprob}
\lefteqn{P_{\nu_{\alpha} \to \nu_{\beta}} \;=} \nonumber\\
&&-4 \sum_{k>j} Re\{U_{\alpha k}^*U_{\beta k}U_{\alpha j} U_{\beta j}^*\}\sin^2\left(\frac{\Delta m^2_{kj} c^4}{4\hbar c }\frac{L}{E}\right)\nonumber\\
&&+ 2\sum_{k>j} Im\{U_{\alpha k}^*U_{\beta k}U_{\alpha j} U_{\beta j}^*\}\sin\left(\frac{\Delta m^2_{kj} c^4}{2\hbar c }\frac{L}{E}\right)\;,
\end{eqnarray}

where $\Delta m^2_{kj}= m^2_k -m^2_j$. As in the neutrino oscillation experiments,
the known quantity is the distance $L$ between the source and the detector and not the propagation time $t$, therefore
the propagation time $t$ is replaced by the source and detector distance  $L$ in the above equation. This is a valid approximation as
all detected neutrinos in the oscillation experiments are ultrarelativistic.

The allowed ranges of the six oscillation parameters, three mixing angles and three mass squared differences, is obtained by
a global fit to solar, atmospheric, reactor and accelerator neutrino data within the framework of three flavor neutrino oscillations.
For normal ordering, the best fit values of three flavor oscillation parameters are \cite{nfit}
\begin{equation}
\theta_{12} = 33.48^{\circ}\,, \quad \theta_{23} = 42.3^{\circ}\,,\quad \theta_{13} = 8.50^{\circ}\,,
\end{equation}
\begin{equation}
\frac{\Delta m^2_{21} c^4}{10^{-5}\, {\rm eV^2}} = 7.50  \,,\quad \frac{\Delta m^2_{31}(\simeq \Delta m^2_{32})c^4}{10^{-3} \,{\rm eV^2}}  = 2.457  \,.
\end{equation}

Following Ref.~\cite{neutri1} we introduce the occupation number of neutrinos by making the
following correspondence:
\begin{eqnarray}
\ket{\nu_e}    &\equiv&    \ket{1}_e   \otimes    \ket{0}_{\mu}\otimes
\ket{0}_{\tau} \equiv \ket{100}\;, \nonumber \\ \ket{\nu_{\mu}} &\equiv& \ket{0}_e
\otimes    \ket{1}_{\mu}\otimes     \ket{0}_{\tau}\equiv    \ket{010}\;,
\nonumber\\  \ket{\nu_{\tau}} &\equiv&  \ket{0}_e \otimes  \ket{0}_{\mu}\otimes
\ket{1}_{\tau}\equiv \ket{001}\;.
\label{eq:rep}
\end{eqnarray}
Consequently, we can view the time evolution of a flavor eigenstate $\alpha=e,\mu,\tau$ as a three-qubit state, i.e.,
\begin{equation}\label{wstate}
|\Psi(t)\rangle_\alpha =
a_{\alpha e}(t)\;\ket{100}+a_{\alpha \mu}(t)\;\ket{010}+a_{\alpha \tau}(t)\;\ket{001}\;.
\end{equation}
Therefore, flavor oscillations can be related to
time variation of the tripartite
entanglement of single particle states.

\section{Study of quantum information theoretic properties in neutrino oscillations}\label{sec:2}

\begin{figure}[t]
\begin{center}
\includegraphics[width=0.5\textwidth]{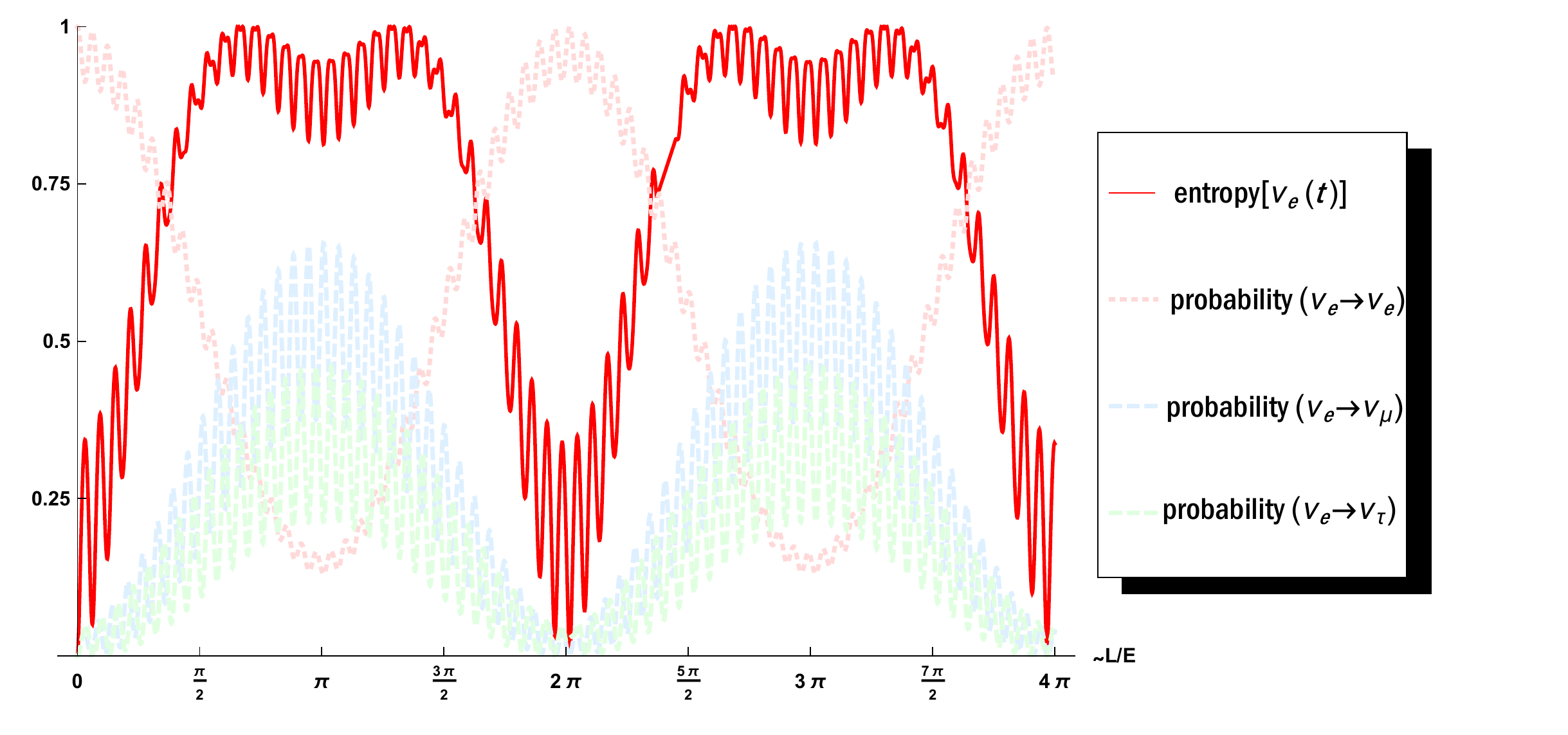}
\caption{Plot of the (normalized) flavor entropy (solid line, red) and the three probabilities ($\nu_e\to\nu_e {\rm(pink,\, dashed)}$ (survival probability, Eq.~(\ref{surviveprob})), $\nu_e\to\nu_\mu {\rm(light blue,\, dashed)}$, $\nu_e\to\nu_\tau {\rm(light green,\, dashed)}$ (oscillation probabilities, Eq.~(\ref{oscprob})) for an initial electron-neutrino state $|\Psi(t=0)\rangle_e=\nu_e(0)$ as a function of the distance traveled per energy $L/E$.}\label{fig:1}
\end{center}
\end{figure}
Separability  or  the lack  of  separability,  i.e., entanglement,  is
defined for a given state according to its possible factorization with
respect to a given algebra  \cite{BTN}. The separability problem is in
general  a  NP-hard problem,  and  only  necessary but  not  generally
sufficient  criteria exists  to  detect  entanglement.  For  bipartite
entangled  quantum systems  it suffices  to ask  whether the  state is
entangled  or not.  In  the  multi-partite case  the  problem is  more
involved  since  there  exist different  hierarchies  of  separability
(defined  later).  We  have  defined the  algebra  by introducing  the
occupation  number of  the three  flavours and  our first  goal is  to
understand  the time  evolution of  neutrino oscillation  in terms  of
tools for classifying and detecting different types of entanglement.

The next  step would  be to take  potential measurement  settings into
account  and  analyse the  different  facets  of correlations  in  the
dynamic of neutrinos.  In particular, we are  interested whether there
are  correlations  stronger  than  those predicted  by  any  classical
theory. The correlations are studied via two different approaches, one
based  on the  dichotomy  between predictions  of  quantum theory  and
different hidden parameter theories, and  the other one quantifies the
various information contents via entropies.

\subsection{Study of the entanglement properties}

Entanglement measures quantify  how much a quantum  state $\rho$ fails
to be separable. Axiomatically, it must be a nonnegative real function
of a state which cannot  increase under local operations and classical
communication (LOCC), and  is zero for separable  states.  An entropic
function generally quantifies the average information gain by learning
about  the outcome  obtained by  measuring a  system. The  von Neumann
entropy, a quantum mechanical analogue  of Shannon entropy, is defined
by $S(\rho)=-\rho \log \rho$ and is zero for pure states and $\log(d)$
for the  totally mixed state, where  $d$ is system dimension,  and the
log function usually  refers to base 2.  The  entanglement content can
be computed by the entropy of  the subsystems since the full system is
pure.

Considering the three possible partial traces of the three-qubit state under investigation, we obtain a concave function of the single-mode probabilities $|a_{\alpha \beta}(t)|^2$, i.e., with $\rho_j:=\textrm{Tr}_{\textrm{all but not subsystem} j}|\Psi(t)\rangle_\alpha\langle \Psi(t)|_\alpha$
\begin{eqnarray}\label{flavorentropy}
\lefteqn{S_{\rm{flavour}}(|\Psi(t)\rangle_\alpha)\;=\;-\sum_{j=e,\mu,\tau} \rho_j \log \rho_j}\nonumber\\
&=& - \sum_{\beta=e,\mu,\tau} |a_{\alpha \beta}(t)|^2 \log |a_{\alpha \beta}(t)|^2\nonumber\\
&&- \sum_{\beta=e,\mu,\tau} (1-|a_{\alpha \beta}(t)|^2) \log (1-|a_{\alpha \beta}(t)|^2),
\end{eqnarray}
which  we  call the  flavor  entropy.   This  function is  plotted  in
Fig.~\ref{fig:1}  together with  the survival  probabilities $\nu_e\to
\nu_e$, Eq.~(\ref{surviveprob}), and the oscillation probabilities $\nu_e\to \nu_\mu,\,\nu_e\to
\nu_\tau$, Eq.~(\ref{oscprob}).

Since          the         flavor          entropy
$S_{\rm{flavour}}(|\Psi(t)\rangle_e)$ is  nonzero for almost  all time
instances,  the state  is entangled.  For this  and all  the following
plots, we  use as units,  the oscillation period  of an electron  to a
muon-neutrino. Since  there are tiny  changes in the behaviour  in one
period due to  the existence of the third flavour,  we always plot two
periods. When the amount of all three probabilities, both the survival
as well  as the  oscillation probabilities,  become nearly  equal, the
flavour  entropy becomes  maximal. Then  the $\nu_\mu$  and $\nu_\tau$
oscillation probabilities become greater than the survival probability
of $\nu_e$  resulting in a  decrease in  the uncertainty of  the total
state   followed  by   an   increase,  when   the  probabilities   get
closer. Next, the  uncertainty in the total state drops  again and the
pattern gets repeated.

\begin{figure*}
\begin{center}
(a)\includegraphics[width=0.3\textwidth]{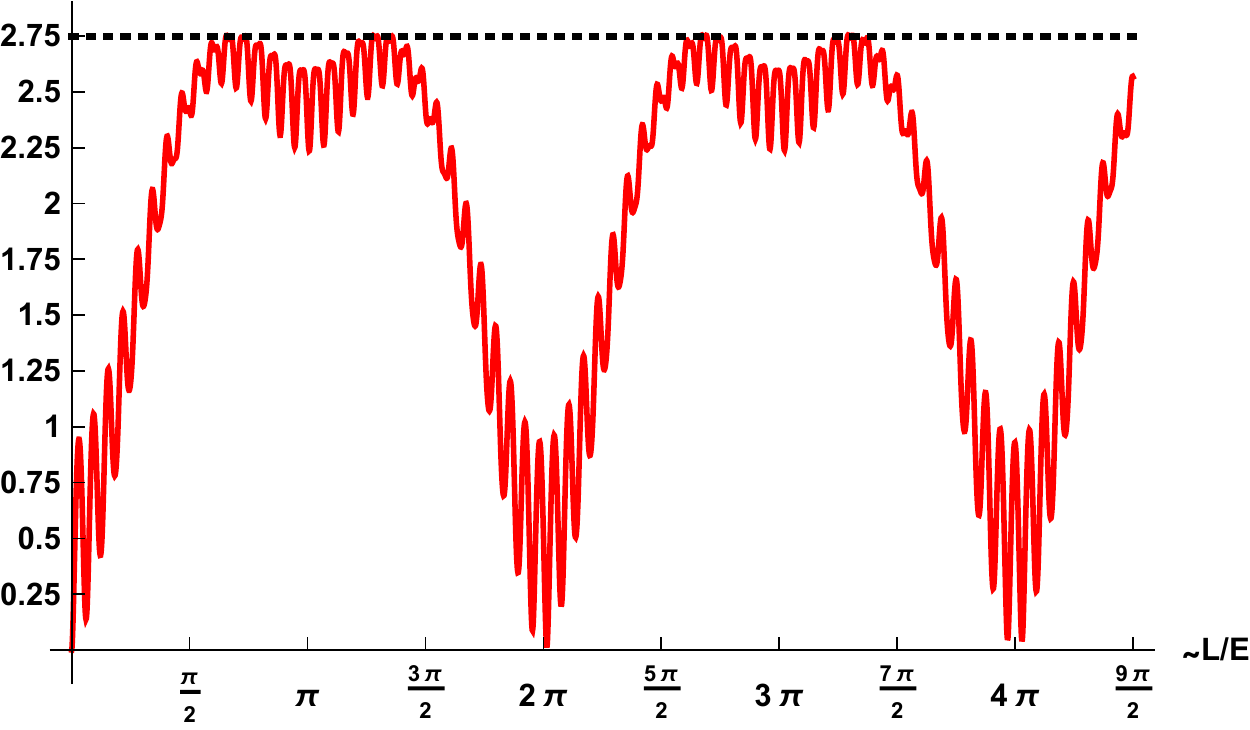}
(b)\includegraphics[width=0.3\textwidth]{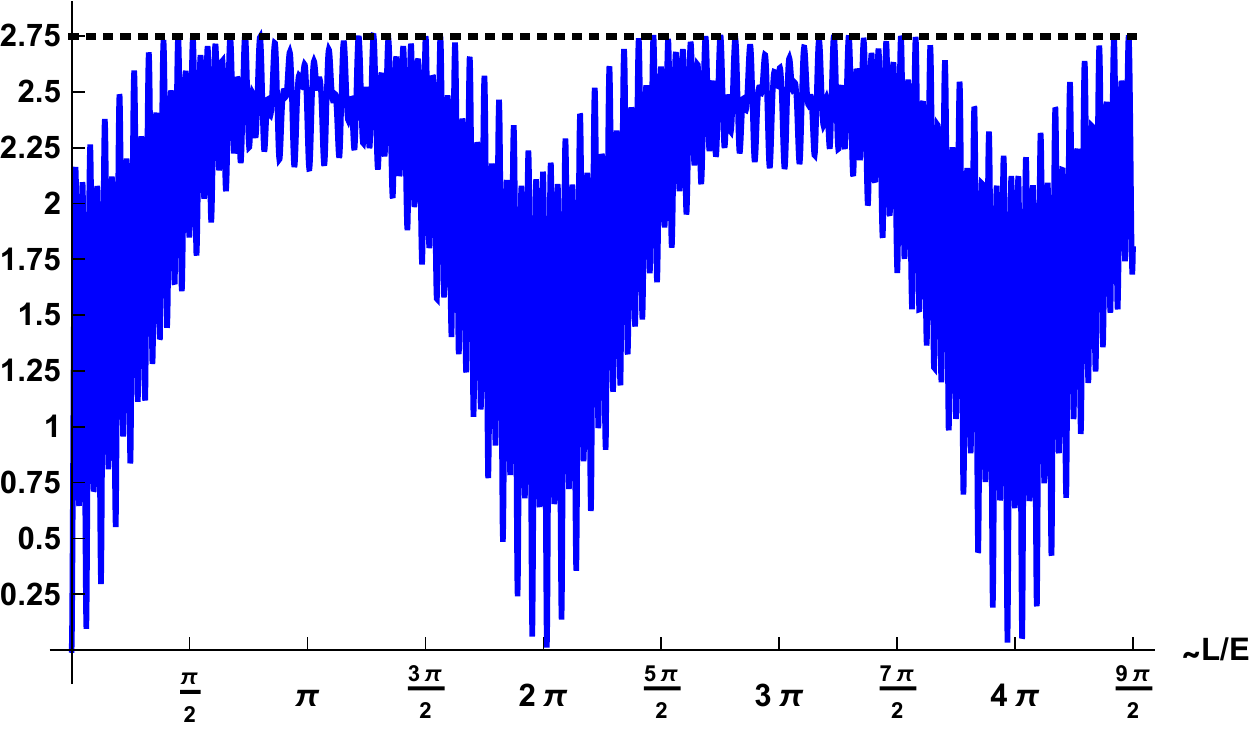}
(c)\includegraphics[width=0.3\textwidth]{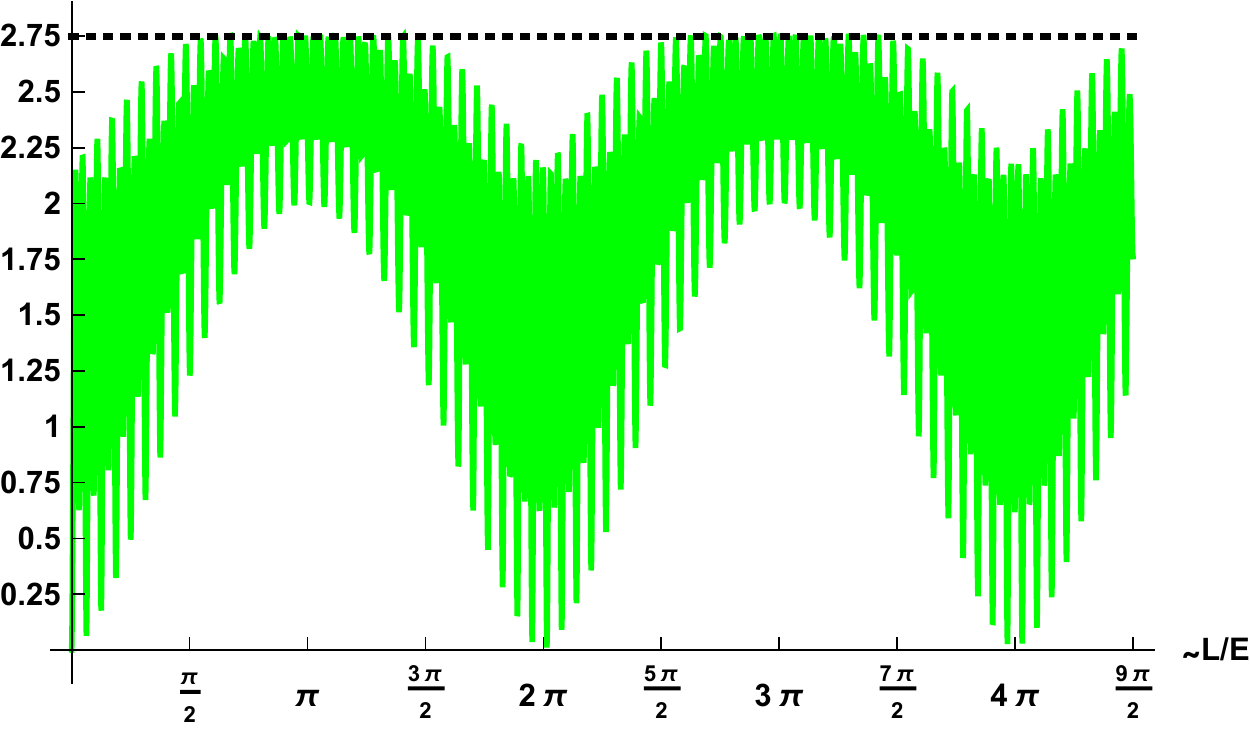}
\caption{Plot the (not-normalized) flavour entropy $S_\textrm{flavor}$, Eq.~(\ref{flavorentropy}), for the three initial flavour states (a) $\nu_{e}$, (b) $\nu_{\mu}$, (c)$\nu_{\tau}$ in terms of the distance traveled per energy $L/E$ in units of the oscillation period of the two lightest neutrinos. The horizontal (dotted) line corresponds to the value of the $W$-state.}\label{fig:entanglemententropy}
\end{center}
\end{figure*}

The entropy of all the three possible flavour entropies are compared in Fig.~\ref{fig:entanglemententropy} showing that for the muon- and tau-neutrinos the entropy is non-zero for almost all time instances. Compared to the electron-neutrino evolution, the flavor uncertainty of the  other two flavours oscillates more rapidly and with higher amplitudes,  reaching the maximal value more often.

Let us now refine the picture by investigating the type of entanglement in neutrino oscillations. A tripartite pure state can be, for example written as
\begin{eqnarray}
|\psi_{k=3}\rangle&=&|\phi_A\rangle\otimes|\phi_B\rangle\otimes|\phi_C\rangle,\nonumber\\
|\psi_{k=2}\rangle&=&|\phi_A\rangle\otimes|\phi_{BC}\rangle,\quad|\phi_B\rangle\otimes|\phi_{AC}\rangle,\nonumber\\
&&\textrm{or}\quad |\phi_{AB}\rangle\otimes|\phi_{C}\rangle,\nonumber\\
|\psi_{k=1}\rangle &=&|\psi\rangle_{ABC},
\end{eqnarray}
where $k$ gives the number of partitions dubbed the $k$-separability. If $k$ equals the number of involved states, in our case $k=3$, the joint state is called fully separable, else it is partially separable. An important class of states are those that are not separable within any bi-partition, and are called \textit{genuinely multi-partite entangled}. In general they allow for applications that outperform their classical counterparts, such as secret sharing \cite{HBB,SHH}. It should be noted that since a $k=3$-separable state is necessarily also $k=2$-separable, $k$-separable states have a nested-convex structure.

Among the genuinely multi-partite entangled states, there are two subclasses known for three qubit states, the GHZ- and W-type of states. In Ref.~\cite{HMGH} a general framework was introduced to detect and define different relevant multipartite entanglement subclasses and refined in several follow ups. In particular it has been shown to allow for a self-consistent classification also in a relativistic framework~\cite{lorentzmeasure}. Generally, one would expect from a proper classification of different types of entanglement that for a relativistically boosted observer, which causes a change of the observed state, but not of the expectation value, that it remains in a certain entanglement class. We will therefore investigate this Lorentz invariant criterion, though let us emphasise that we do not take any relativistic effects of a boosted observer into account in this contribution.

The necessary criterion for a tripartite qubit state with one excitation (``$1$'')  to be bipartite reads
\begin{widetext}
\begin{eqnarray}\label{Dicketripartite}
&&Q_{Dicke}^1(\rho)\;=
2 \left|\langle 001|\rho|010\rangle\right|+2 \left|\langle 001|\rho|100\rangle\right|+2 \left|\langle 010|\rho|100\rangle\right|
-\biggl(\langle 001|\rho|001\rangle+\langle 010|\rho|010\rangle+\langle 100|\rho|100\rangle\nonumber\\
 &&\quad+ 2 \sqrt{\langle 000|\rho|000\rangle\cdot \langle 011|\rho|011\rangle}+2 \sqrt{\langle 000|\rho|000\rangle\cdot \langle 101|\rho|101\rangle}
 +2 \sqrt{\langle 000|\rho|000\rangle\cdot \langle 110|\rho|110\rangle}\biggr)\;\leq 0.
\end{eqnarray}
\end{widetext}
If  this criterion  is  violated  the state  $\rho$  has no  bipartite
decompositions,  i.e., it  is  genuinely  multipartite entangled.  The
positive terms are exactly the only non-zero off-diagonal terms of the
$W$-state, $|W\rangle = \frac{1}{\sqrt{3}}\lbrace |100\rangle+|010\rangle+|001\rangle\rbrace$,
with one excitation in the computational basis,
whereas the  negative terms are  only diagonal terms. Note  that these
negative terms are  all zero for the  $W$-state in  the given basis  such that
only this state obtains the maximum value.

\begin{figure*}
\begin{center}
(a)\includegraphics[width=0.3\textwidth]{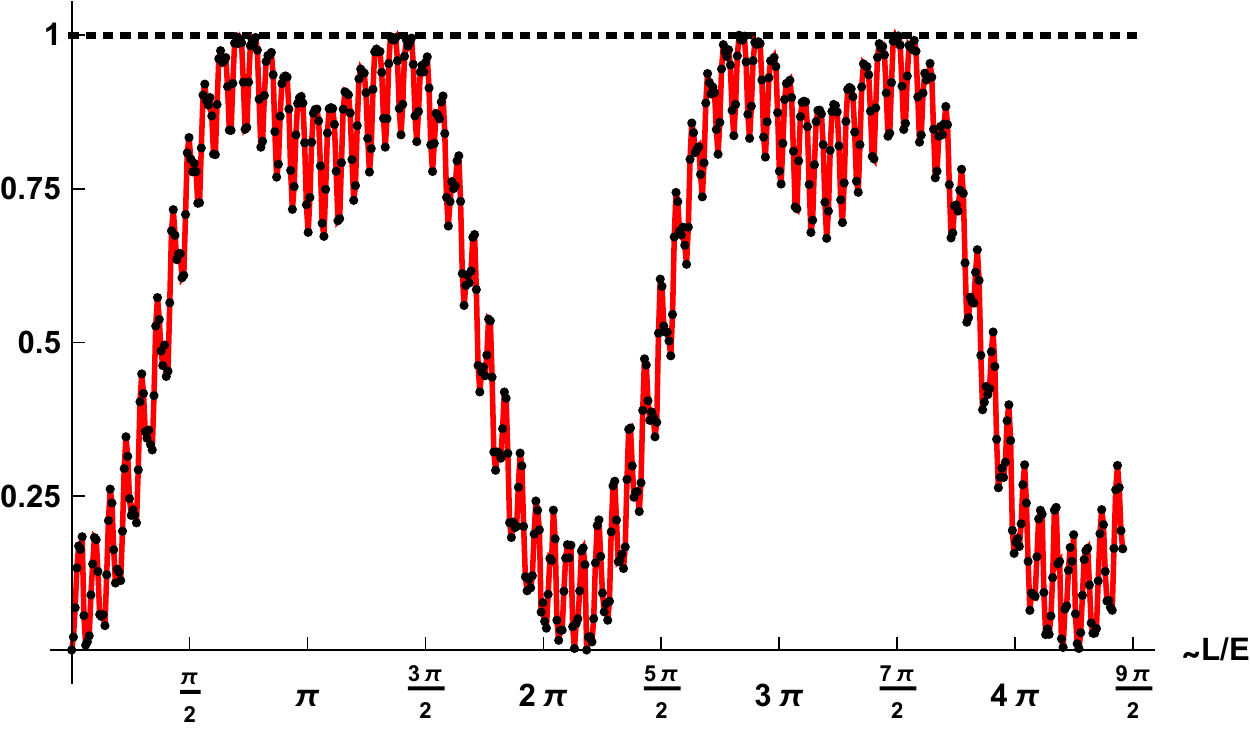}
(b)\includegraphics[width=0.3\textwidth]{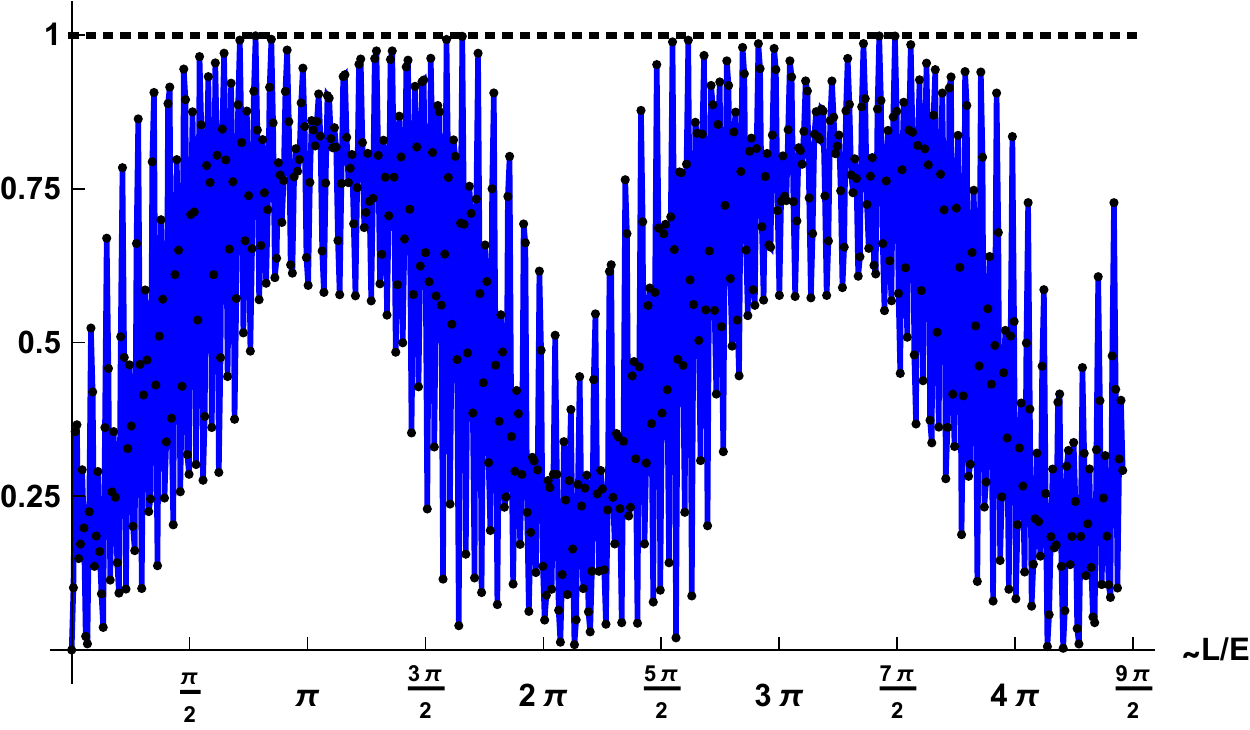}
(c)\includegraphics[width=0.3\textwidth]{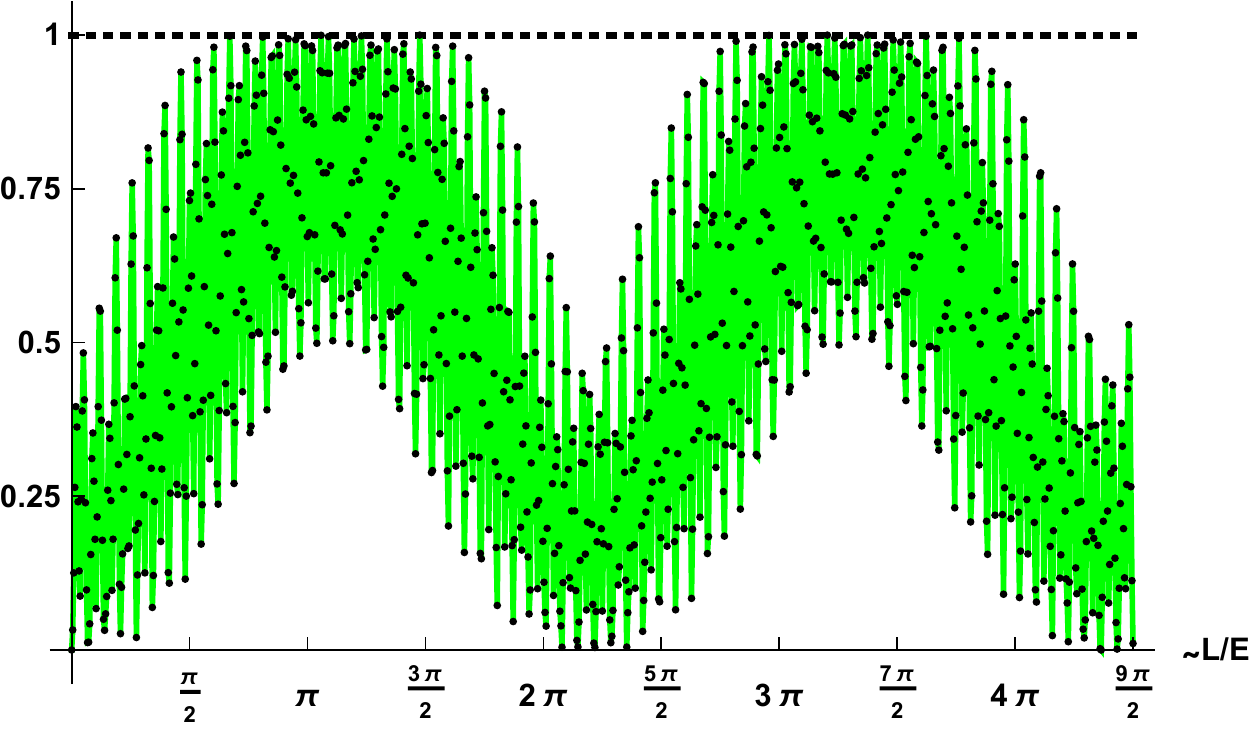}
\caption{Plot of the criterion $Q_{Dicke}^1$ detecting genuine multipartite entanglement, Eq.~(\ref{Dicketripartite}), optimized over local unities for the three initial flavour states (a) $\nu_{e}$, (b) $\nu_{\mu}$, (c)$\nu_{\tau}$ with respect to the distance traveled per energy $L/E$ in units of the oscillation period of the two lightest neutrinos ($300$ data points). The criterion detects genuine multipartite entanglement if greater than zero and is maximal ($=1$) only for the $W$-state.}\label{fig:qmopt}
\end{center}
\end{figure*}

Obviously, this criterion  depends on the basis  representation of the
state $\rho$ and has therefore to be optimized over all local unitary
operations. Indeed, taking the  ``flavour basis'' as the computational
basis, Eq.~(\ref{wstate}), the unoptimized criterion becomes
\begin{eqnarray}
&&2 |a_{\alpha e}(t)a_{\alpha \mu}(t)|+2 |a_{\alpha e}(t)a_{\alpha \tau}(t)|+ 2 |a_{\alpha \mu}(t)a_{\alpha \tau}(t)|\nonumber\\
&&-\underbrace{(|a_{\alpha e}(t)|^2+|a_{\alpha \mu}(t)|^2+|a_{\alpha \tau}(t)|^2)}_{=1}\quad \leq 0\;,
\end{eqnarray}
which is not  violated for all times.  Consequently, optimization over all local unitaries has to be taken into account for
each time point  and is plotted for an initial  electron-, muon- and tau-neutrino in Fig.~\ref{fig:qmopt}. We find that the states at each time
point  are  always genuine  multipartite  entangled  if at  least  two
amplitudes of  the state, Eq.~(\ref{wstate}), are  non-zero, i.e., for
almost all  time instances.
The results  depicted in Fig.~\ref{fig:qmopt}  also prove that  in the
course of the  time evolution the genuine multipartite  $W$ state (all
amplitudes equal  to $\frac{1}{\sqrt{3}}$)  is reached.  Hence, Nature
exploits  the   maximum  genuine  multipartite  entanglement   in  the
occupation number basis.

\subsection{Genuine multipartite nonlocality} \label{sec:3loc}

\begin{figure*}
\begin{center}
(a)\includegraphics[width=0.3\textwidth]{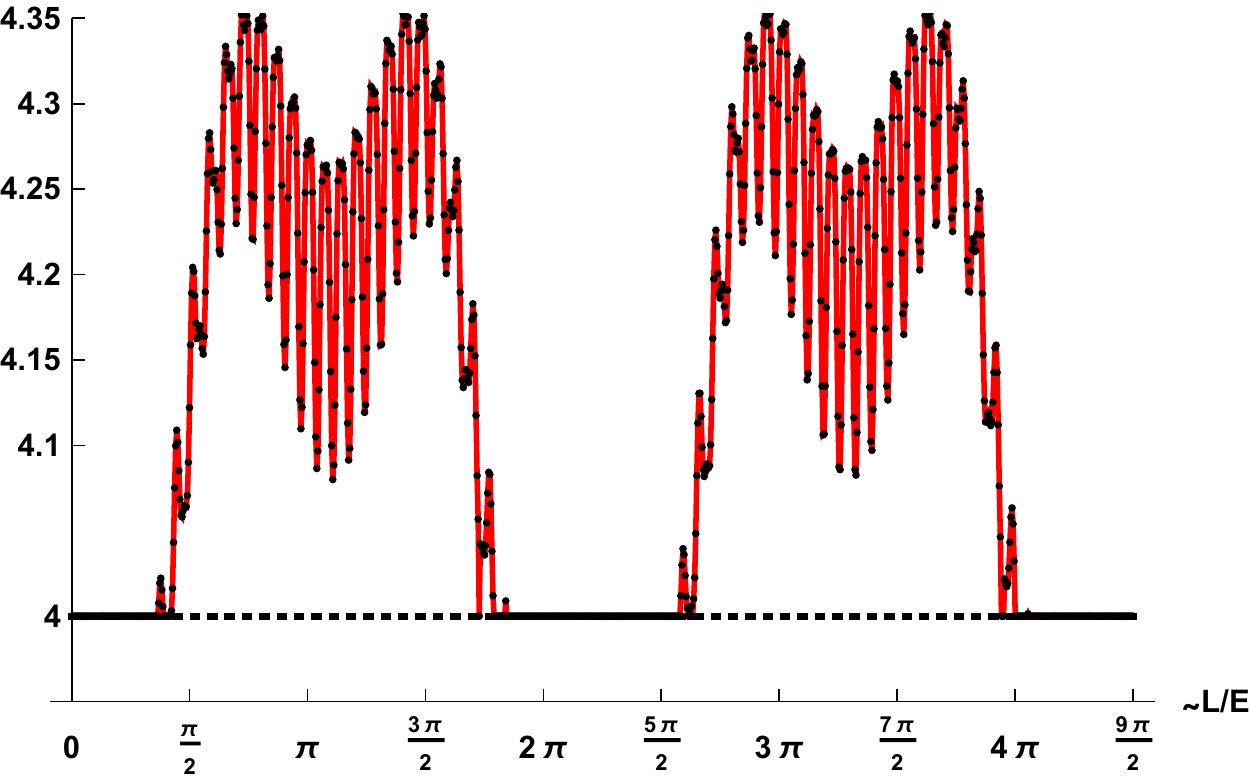}
(b)\includegraphics[width=0.3\textwidth]{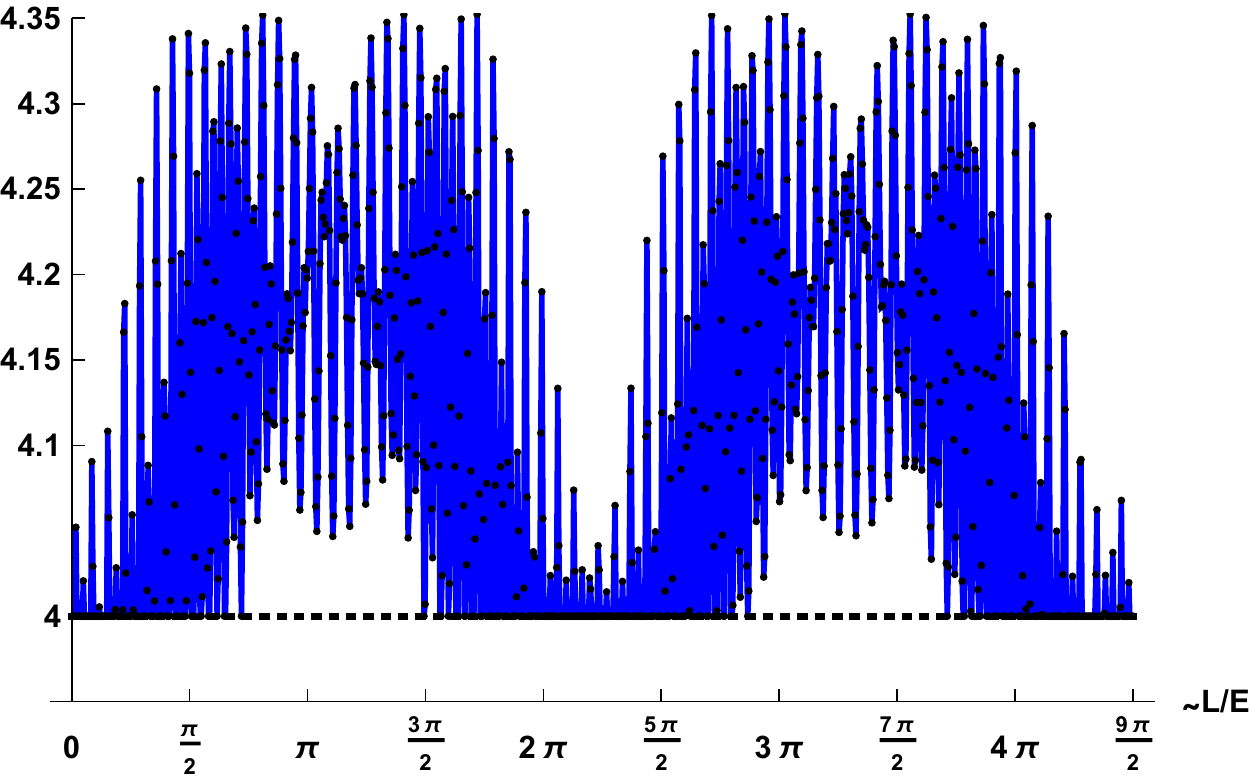}
(c)\includegraphics[width=0.3\textwidth]{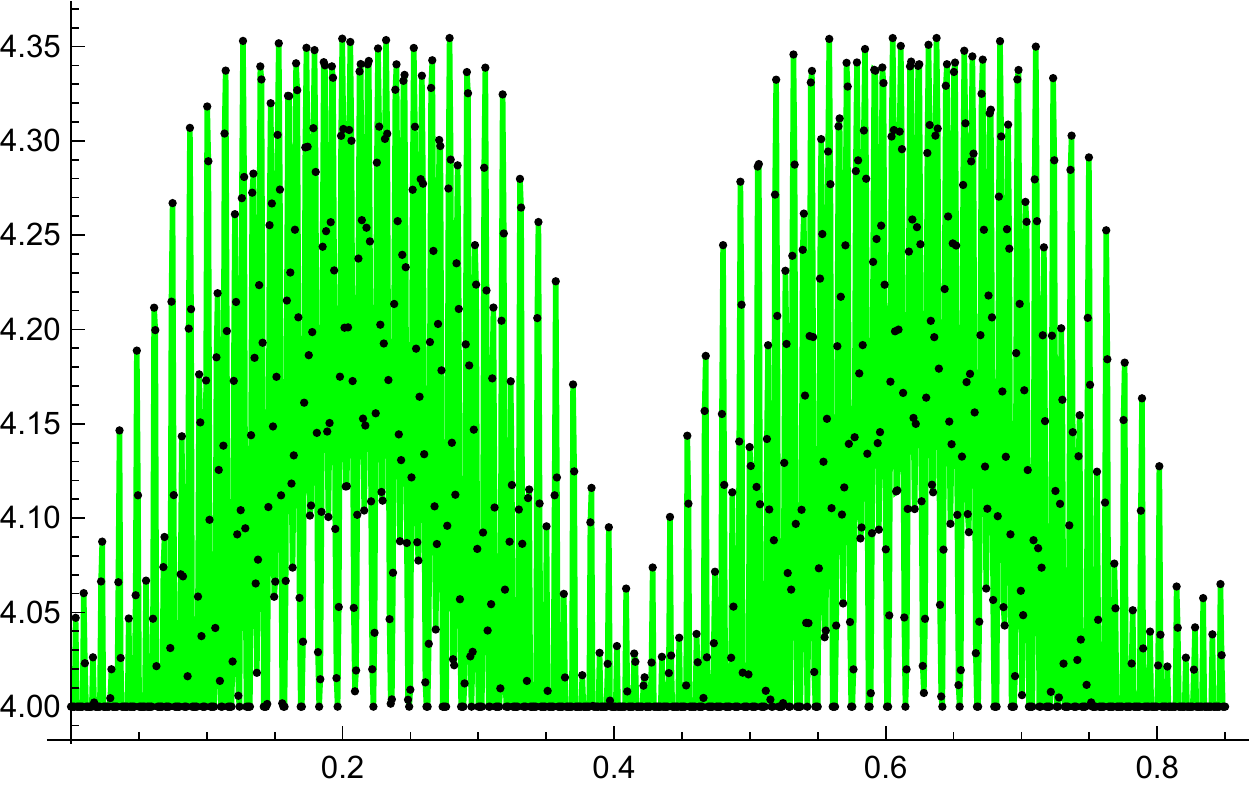}
\caption{Plot of the Svetlichny criteria detecting genuine multipartite nonlocality, Eq.~(\ref{SvetlichnyCrit}), optimized over possible bipartitions and optimized over all six different observables for the three initial flavour states (a) $\nu_{e}$, (b) $\nu_{\mu}$, (c)$\nu_{\tau}$ as a function of the distance traveled per energy $L/E$ in units of the oscillation period of the two lightest neutrinos ($300$ data points). The criterion detects genuine multipartite nonlocality if the value is above $4$.}\label{fig:sopt}
\end{center}
\end{figure*}

We now ask the question whether in the course of the flavor oscillations, Bell-type nonlocality is persistent, i.e., there are correlations stronger than those predicted by any classical hidden variable theory. For that we investigate the Svetlichny inequalities~\cite{Svet87} which are a sufficient criteria for proving \textit{genuine} tripartite nonlocality. In short, the idea is whether by measuring three observables $A,B,C$  and obtaining the results  $a,b,c$, the probability $P(a,b,c)$ can be assumed to be factorizable as
\begin{eqnarray}
P(a, b, c)&=&\int \mathsf{f}(a b|\lambda)\cdot \mathsf{h}(c|\lambda)\; d\omega(\lambda)\;,
\end{eqnarray}
where  $\mathsf{f},\mathsf{h}$ are  probabilities  conditioned to  the
hidden variable $\lambda$ with  the probability measure $d\omega$. The
factorization, here  chosen between  the partitions $A,B$  versus $C$,
corresponds to  Bell's locality assumption in  his original derivation
if  considered for  two systems.   (The requirement  of a  full
  factorization, i.e., the additional factorization  $\mathsf{f}(a
  b |\lambda)= \mathsf{q}(a\lambda)\cdot \mathsf{r}(b \lambda)$, which
  correspond  to   absolute  locality,   is  explored  later   by  the
  inequalities~(\ref{eq:mermin})).  Then  the necessary  criteria for
such a factorization  of the conditioned probabilities,  qubit $1$ and
$2$ versus $3$, are given by
\begin{eqnarray}\label{SvetlichnyCrit}
\lefteqn{ I^a_{12|3}(\rho)\;=}\nonumber\\
&& Tr\left((ADC+AD'C'+A'D' C-A'D C')\rho\right)\leq 4\;,\nonumber\\
\lefteqn{I^b_{12|3}(\rho)\;=}\nonumber\\
&& Tr\left((AD'C+AD'C'-A'D C'-A'D'C')\rho\right)\leq 4\;,
\end{eqnarray}
with $D=B+B'$  and $D'=B-B'$. Since we are not interested in a particular hidden parameter model, we also consider the two other bipartitions, namely $2|13$ and $3|12$, and take the maximum over them. In Fig.~\ref{fig:sopt}  we have plotted
the maximum  of $I^a$ and $I^b$, over all bipartitions,  for the time evolution  of an initial
electron-, muon- and  tau-neutrino.  In addition, each data point  corresponds to the
maximum  of the  optimization  over all  possible observables  $A,B,C$.
In the  case of an  initial electron-neutrino we find regions in the time evolution when the criterion does
not detect genuine  tripartite nonlocality, whereas for  the two other
neutrino flavours we observe  a stronger oscillating behaviour.
  Summing up, whereas genuine  nonlocal correlation is largely present
  in the time evolution, there are specific time regions when it vanishes.

\begin{figure*}
\begin{center}
(a)\includegraphics[width=0.3\textwidth]{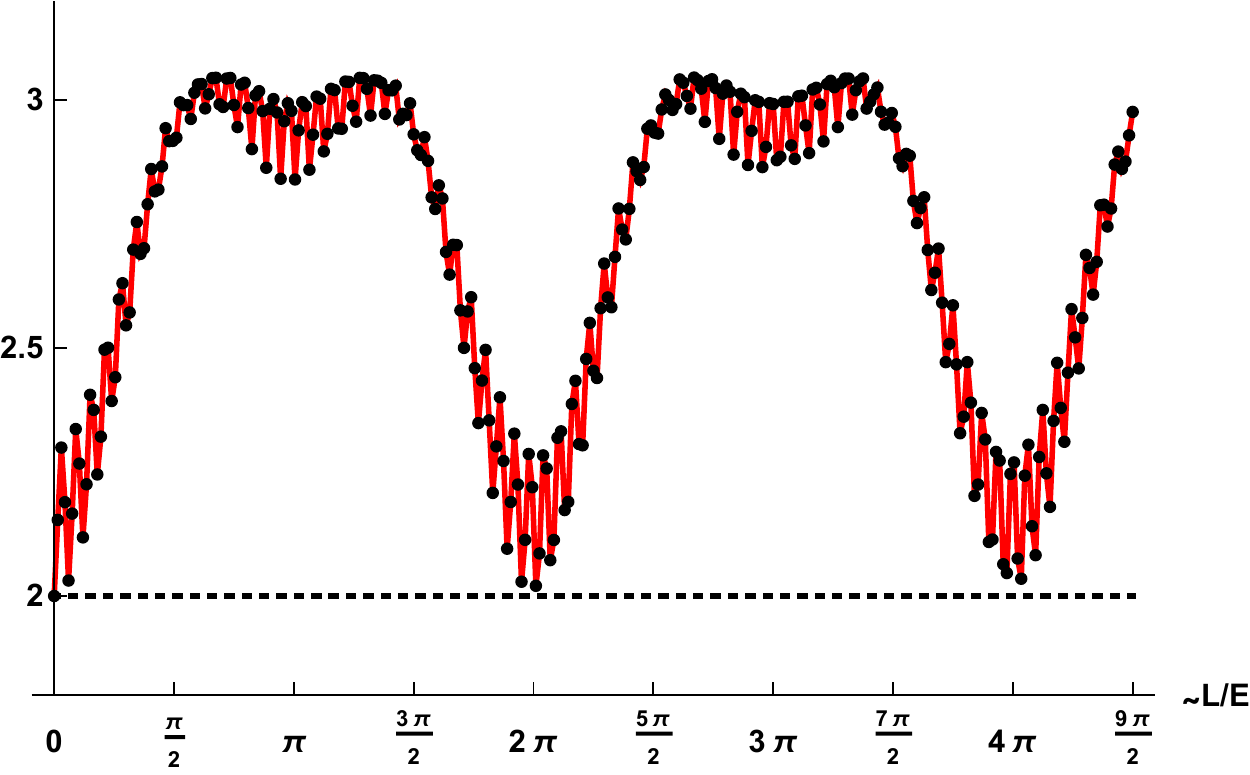}
(b)\includegraphics[width=0.3\textwidth]{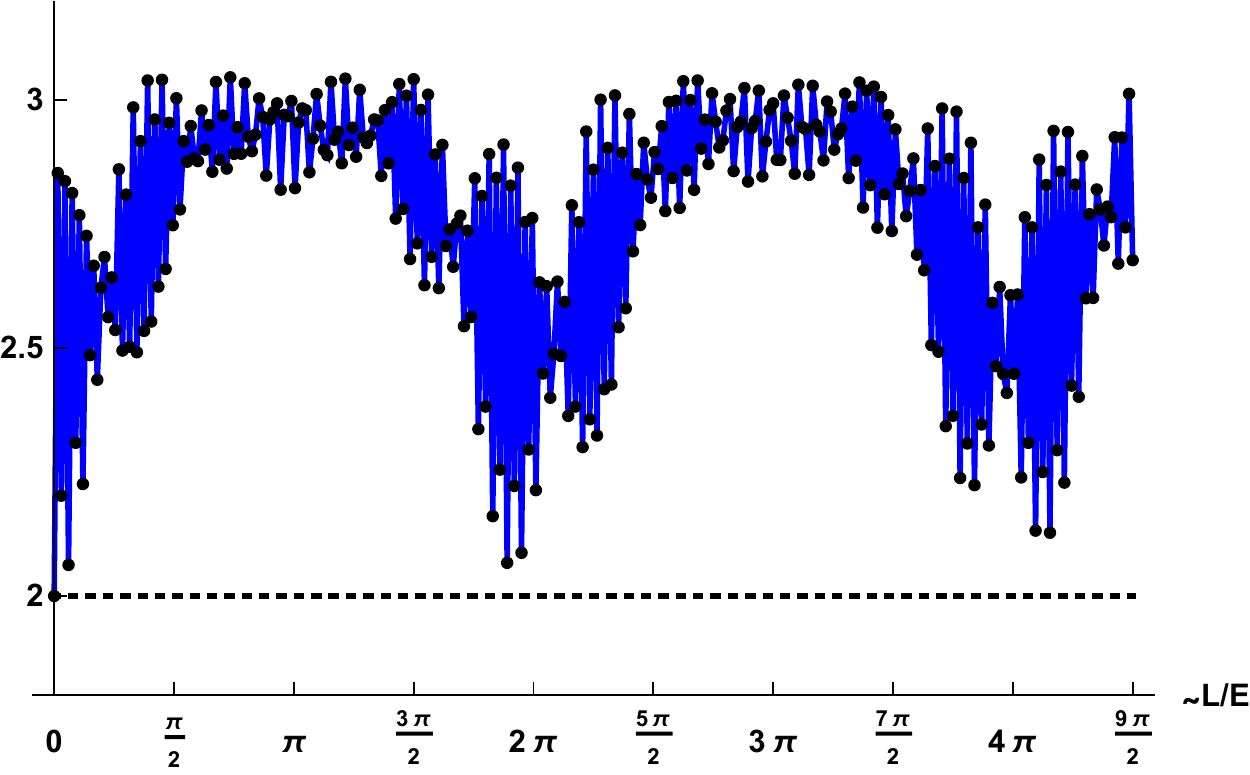}
(c)\includegraphics[width=0.3\textwidth]{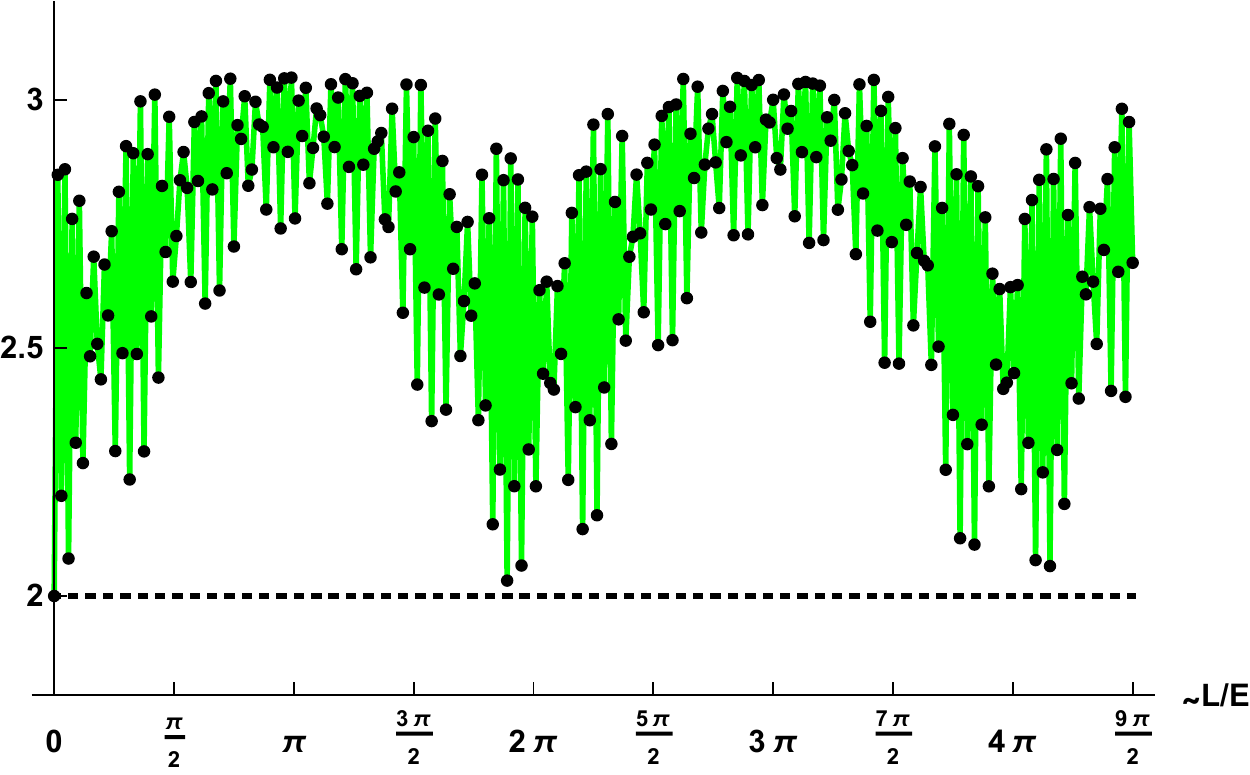}
\caption{Plot of the maximum of $M^a$  and $M^b$, Eq.~(\ref{eq:mermin}), optimized over all involved operators for the three initial flavour states (a) $\nu_{e}$, (b) $\nu_{\mu}$, (c) $\nu_{\tau}$ in terms of the distance traveled per energy $L/E$ in units of the oscillation period of the two lightest neutrinos ($300$ data points). The criterion is above two if and only if no hidden variable model exists.}\label{fig:mermin}
\end{center}
\end{figure*}

Requiring that for all three measurements a hidden parameter model should exist can be revealed by the following set of inequalities~\cite{M90}
\begin{eqnarray}\label{eq:mermin}
M^a(\rho)&=& Tr\left((ADC+AD'C)\rho\right)\leq 2\;,\nonumber\\
M^b(\rho)&=& Tr\left((A'D'C-A'D C')\rho\right)\leq 2\;,
\end{eqnarray}
which are connected to the Svetlichny inequality by $I^a_{12|3}=M^a+M^b$ (see Refs.~\cite{Svet87,ajoy}).  These are the Mermin inequalities and their violation is an indicator of absolute nonlocality. Again we are interested in finding a contradiction to any hidden parameter model, thus we consider all bipartitions and take the maximum. The results are plotted in Fig.~\ref{fig:mermin} (including an optimization over all four arbitrary operators $A,D,C,D'$). For all times (except when the state is separable) the two inequalities are violated when optimized over all measurement settings. This shows that assuming that the mode correlations can be simulated by an ensemble where all three subsystems are correlated to each other for all time instances is not possible. In contrast, correlations simulated by a hybrid nonlocal-local ensemble, captured by inequalities~(\ref{SvetlichnyCrit}), may exist for time instances close to the separable state, however, only for the electron-neutrino dynamics.

To sum up, except for small time regions neutrino oscillations exhibit
all the  strong correlations, entanglement and  Bell-type nonlocality,
that  are considered  to  give  an advantage  to  quantum theory  over
classical  theories  for a  number  of  information processing  tasks.
 For  completeness,  in  the next  section  we  investigate  the
  behavior of a measure of nonclassicality weaker than entanglement.

\subsection{Dissension-a measure of nonclassicality}

Classical mutual information, quantifying the information between two random variables $A$ and $B$, can be defined by
$I(A{:}B)  =  H(A)  -  H(A|B)$, where $H(A)=-\sum_i p_i \log p_i$ is the Shannon entropy of the probabilities $p$ of the outcomes of $A$ and $H(A|B):=H(A)-H(A,B)$   represents  the classical   conditional  entropy and $H(A,B)$ is the joint entropy of the pair of random variables $(A,B)$ (see, e.g., Ref.~\cite{nc}). Mutual information can  be  generalized  for
three random variables $A,B,C$  by any  of the following three equivalent expressions~\cite{CT91}
\begin{widetext}
\begin{eqnarray}
I_1(A{:}B{:}C) &=& H(A,B) - H(B|A) - H(A|B) 
-H(A|C) - H(B|C)
+H(A,B|C), \nonumber \\
I_2(A{:}B{:}C)&=& H(A) + H(B) + H(C) 
-H(A,B)-H(A,C)-H(B,C) + H(A,B,C), \nonumber\\
I_3(A{:}B{:}C)&=&  H(A) + H(B) + H(C) - H(A,B) - H(A,C)
+H(A|B,C)\;.
\label{eq:classical3}
\end{eqnarray}
\end{widetext}
While   the  second  of  these   expressions  suggests  a straightforward  quantum  generalization,  by  replacing  the  Shannon
entropy  by  the corresponding  von  Neumann  entropy $S(\rho)  \equiv
-\textrm{Tr}(\rho\log\rho)$,  the  first   and  third  expressions
lead to  complications since the
average conditioned  entropy depends on the  basis chosen  and
on the choice of the random variables $A,B,C$.  Let us point out that in strong contrast to the bipartite mutual information the tripartite mutual information may  also be negative. This is the case if for instance knowing the random variable $C$ enhances the correlation between $A$ and $B$. Following the concept of quantum discord \cite{olivierzurek,henderson}, which quantifies nonclassical correlations, in Ref.~\cite{chakrabarty} two measures for nonclassicality, called dissension, were introduced:
\begin{eqnarray}
D_1(A{:}B{:}C) &=& J_1(A{:}B{:}C) - J_2(A{:}B{:}C), \nonumber \\
D_2(A{:}B{:}C) &=& J_3(A{:}B{:}C) - J_2(A{:}B{:}C)\;,
\label{eq:D1D2}
\end{eqnarray}
where $J_i$ are the quantum analogs of classical tripartite mutual information $I_i$, Eq.~(\ref{eq:classical3}), namely
 \begin{eqnarray}
J_1(A{:}B{:}C)  &=& S(A,B)  -  S(B|\Pi^A) -  S(A|\Pi^B)  - S(A|\Pi^C) \nonumber \\ &-&
S(B|\Pi^C) + S(A,B|\Pi^C), \nonumber \\
J_2(A{:}B{:}C) &=& S(A) + S(B) +
S(C) \nonumber\\ &-& S(A,B)-S(A,C)-S(B,C) + S(A,B,C), \nonumber\\
J_3(A{:}B{:}C) &=& S(A) + S(B) +
S(C) - S(A,B) - S(A,C) \nonumber\\ &+& S(A|\Pi^{B,C})\;.
\label{eq:quantum3}
\end{eqnarray}
Here    $S(A|\Pi^B)   =    \sum_k   p_k    S(\rho_{A|\Pi^B_k})$   with\\
$\rho_{A|\Pi^B_k} =  (\Pi^B_k \rho_{AB} \Pi^B_k)/p_k$ and  $p_k \equiv
\textrm{Tr}(\Pi^B_k \rho_{AB})$ is the probability that outcome $k$ is
obtained.  It is assumed that the basis  of $\Pi^B_k$ is chosen such as to
minimize  the uncertainty.   A  given  state   is  denoted to  be
nonclassical   for   any   departure    of   $D_1$   or   $D_2$   from
$0$.  Here $D_1$  and $D_2$  deviations from  zero can  be
  associated to nonclassicality accessed  by only one-mode or two-mode
  measurements, respectively.

We find that $J_2$ is always zero, since $S(A,B,C)$ is zero because the total state is pure and since $S(A)=S(B,C),\;S(B)=S(A,C),\;S(C)=S(A,B)$, this being a particularity of the $W$ class of  states. Furthermore, any permutation of the entropy $S(A|\Pi^{B,C})=0$ is zero, since any projection onto the two-mode subspace gives a pure state which has zero uncertainty. Consequently, the relevant measures reduce in our case to
\begin{eqnarray}\label{dissension}
\lefteqn{D_1(\Psi_\alpha)\;=\;\bigl\lbrace}\nonumber\\
&&S(e,\mu)-S(e,\Pi^\mu)-S(e,\Pi^\tau)-S(\mu,\Pi^e)-S(\mu,\Pi^\tau),\nonumber\\
&& S(e,\tau)-S(e,\Pi^\mu)-S(e,\Pi^\tau)-S(\tau,\Pi^e)-S(\tau,\Pi^\mu),\nonumber\\
&&S(\mu,\tau)-S(\mu,\Pi^e)-S(\mu,\Pi^\tau)-S(\tau,\Pi^e)-S(\tau,\Pi^\mu)\bigr\rbrace\;,\nonumber\\
\lefteqn{D_2(\Psi_\alpha)\;=\;\bigl\lbrace S(e),S(\mu),S(\tau)\bigr\rbrace\;.}
\end{eqnarray}
Here the three terms in the bracket of $D_1$ refer to the single electron-neutrino, single muon-neutrino and single tau-neutrino mode measurements, respectively. The three terms in the bracket of $D_2$ refer to joint bipartite measurements in the muon-tau, electron-tau, muon-electron mode subspace, respectively (which are minimized to zero).

\begin{figure*}
\begin{center}
(a)\includegraphics[width=0.45\textwidth,height=0.25\textwidth]{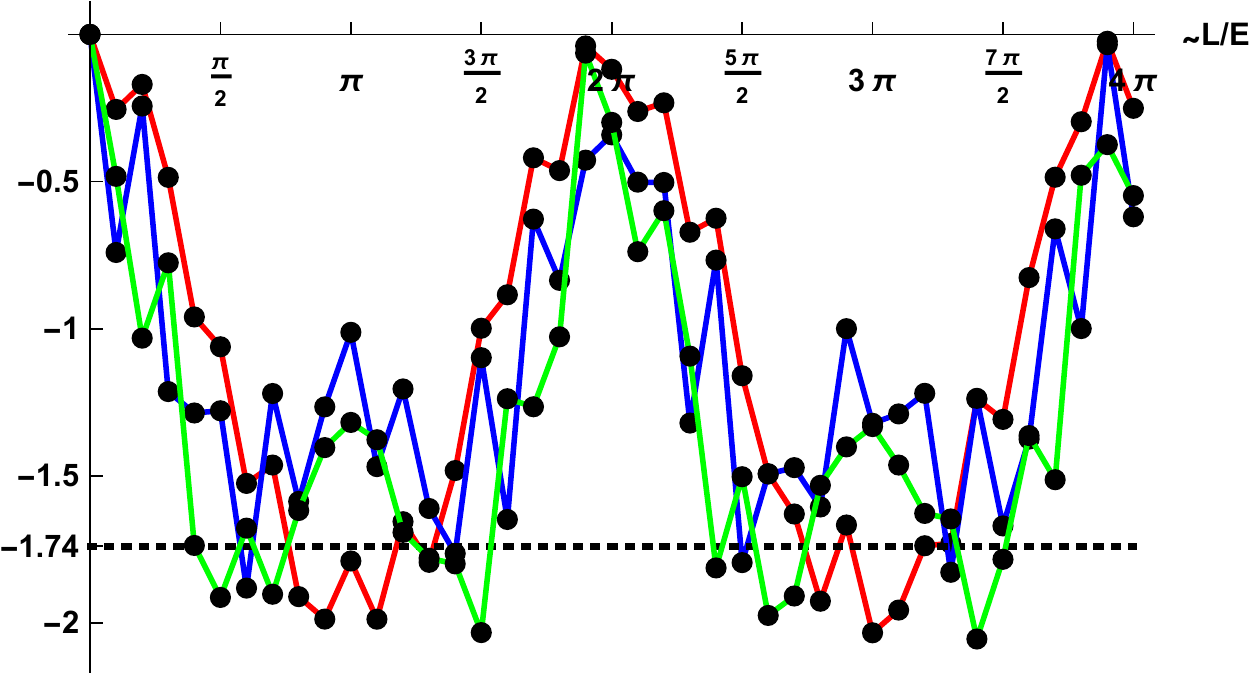}
(b)\includegraphics[width=0.45\textwidth,height=0.25\textwidth]{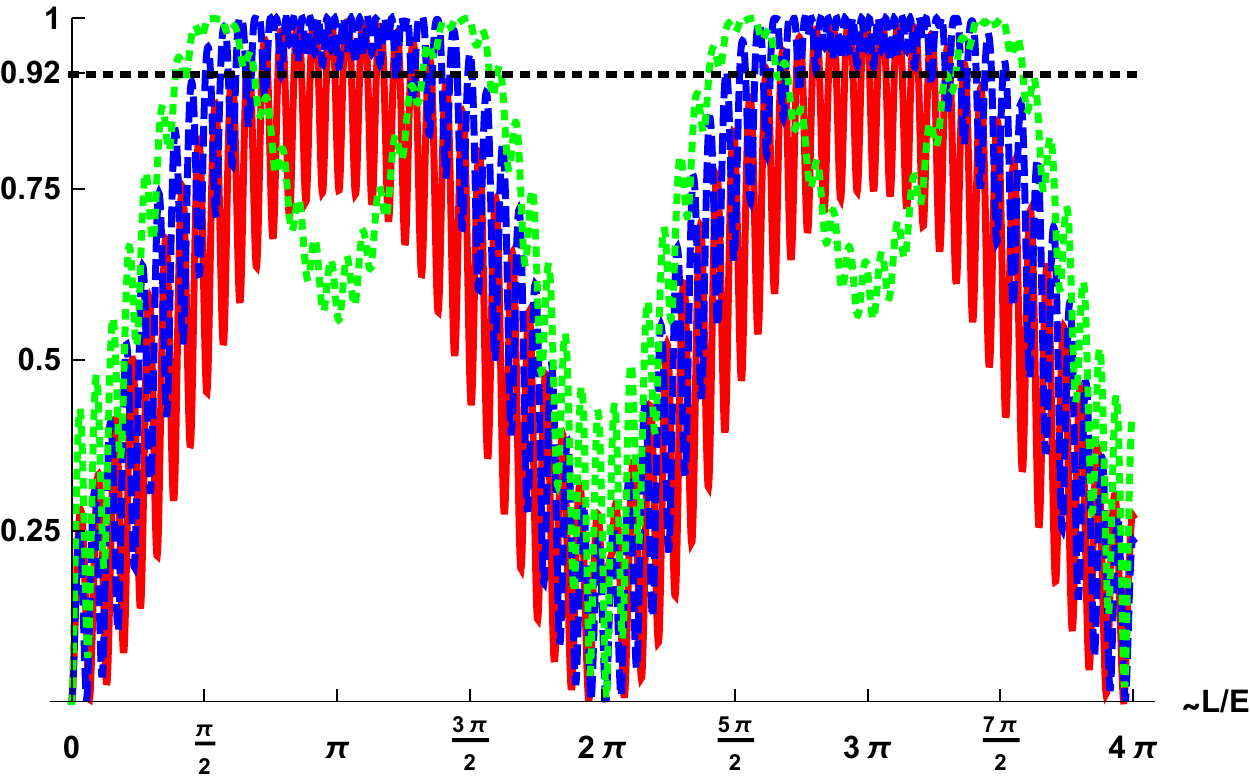}
\caption{Plots         of        the         dissensions,
    Eq.~(\ref{dissension}), minimized over all projective measurements
    for the time evolution of an  initial $\nu_e$ as a function  of the
    distance per  energy $L/E$: (a)  single-mode measure $D_1$  and (b)
    two-mode measure  $D_2$. The colors  encode the dependence  on the
    reference    mode:    (red,     $\nu_e$),    (blue,    $\nu_\mu$),
    (green,$\nu_\tau$).   The  horizontal  lines  corresponds  to  the
    optimized values  of the  $W$-state, respectively.  Curiously, the
    measures    are   always    non-zero,   detecting    non-classical
    correlations,  and  exceed the  value  of  the $W$-state  in  both
    cases.}\label{fig:dissension}
\end{center}
\end{figure*}

In  Fig.~\ref{fig:dissension}   we  plot  the   dissensions  $D_1,D_2$
minimized over all  projective measurements for the  time evolution of
an initial electron-neutrino.   The first notable point is
  that both measures are very sensitive to whether the nonclassicality
  is accessed  by single or  bipartite measurements and  both measures
  are non-zero for almost all  times. Interestingly, we find
  that for both measures  $\min{D_1},\min{D_2}$ and all measurement types there are time regions
  for  which  the  value  exceeds  the  corresponding  value  for  the
  $W$-state, which has $(\min{D_1},\min{D_2})=(-1.738,0.918)$. For single measurements dissension $D_1$ is  still considerably smaller than  the values for
  the   GHZ  state   ($\min{D_1}=-3$), in contrast to $D_2$ where $\min{D_2}=1$.  Moreover, a strong ``twin-humped'' pattern of $D_2$ in the time evolution is found for joint measurements in the subspace of the two heavier neutrinos showing the existence of the  third neutrino flavour ($\tau$).

\section{Conclusions and Outlook}


To sum up,  we have computed several  information theoretic quantities
detecting and  classifying correlations for  the time evolution  of an
initial electron-, muon- or tau-neutrino.  We find that for almost all
time instances the neutrino  states exhibit genuine quantum features.

 We have analysed in detail the dynamics of initial neutrino states via various types of entanglement properties, correlations that cannot be simulated by realistic hidden variable theories and non-classical correlations revealed by mutual information measures. In particular,  dissension turned out to be larger than that for the perfect $W$-state (Dicke state), for some time values, in strong contrast to the measures not involving measurements, i.e., the flavor entropy and the criterion detecting genuine multipartite entanglement. What physical  significance this carries, if any, remains to be seen.

Qualitatively, there are differences between an initial electron-neutrino and the other  two neutrinos, i.e., with the former
  showing   less  nonclassical   features   when compared to  its   heavier
  counterparts,  a point  that may  merit further  scrutiny.  In  any
case, we can  conclude that foundational issues are  more prominent in
accelerator  experiments  (mainly  producing muon-neutrinos)  than  in
reactor experiments (mainly producing electron-neutrinos).

The weak  force, being  one of  the four  known fundamental  forces in
Nature, dominant in the flavour changing process of neutrinos, reveals
strong genuine quantum features such as also shown for weakly decaying
spinless  $K$-mesons~\cite{Hiesmayr:2012} or  for the  weakly decaying
half integer  spin hyperons~\cite{BH_Hyperon}. The next  step would be
to understand how and whether  Nature takes advantage of these strong
quantum  correlations for   information  processing in  a natural
  setting.

\begin{acknowledgements}
B.C. gratefully thanks the Austrian Science Fund (FWF-26783).
\end{acknowledgements}


\begin{thebibliography}{9}
%
%
\bibitem{nc}
M.~A.~Nielsen and I.~L.~Chuang, Quantum Computation and Quantum Information, Cambridge University Press, (2010).

\bibitem{meson1}
  P.~H.~Eberhard,
  Testing the nonlocality of quantum theory in two kaon systems,
  Nucl.\ Phys.\ B {\bf 398}, 155 (1993).
  \bibitem{meson2}
  A.~Di Domenico,
  Testing quantum mechanics in the neutral kaon system at a Phi factory,
  Nucl.\ Phys.\ B {\bf 450}, 293 (1995).
  \bibitem{meson3}
  F.~Uchiyama,
  Generalized Bell's inequality in two neutral kaon systems,
  Phys.\ Lett.\ A {\bf 231}, 295 (1997).
  \bibitem{meson4}
  F.~Selleri,
  Incompatibility between local realism and quantum mechanics for pairs of neutral kaons,
  Phys.\ Rev.\ A {\bf 56}, 3493 (1997).
  \bibitem{meson5}
   B.~Ancochea, A.~Bramon and M.~Nowakowski,
  Bell inequalities for K0 anti-K0 pairs from phi resonance decays,
  Phys.\ Rev.\ D {\bf 60}, 094008 (1999).
  \bibitem{meson6}
    R.~A.~Bertlmann, W.~Grimus and B.~C.~Hiesmayr,
  Bell inequality and CP violation in the neutral kaon system,
  Phys.\ Lett.\ A {\bf 289}, 21 (2001).
  \bibitem{meson7}
    M.~Genovese, C.~Novero and E.~Predazzi,
  Conclusive tests of local realism and pseudoscalar mesons,
  Found.\ Phys.\  {\bf 32}, 589 (2002).
  \bibitem{meson8}
    A.~Bramon and G.~Garbarino,
  Novel Bell's inequalities for entangled K0 anti-K0 pairs,
  Phys.\ Rev.\ Lett.\  {\bf 88}, 040403 (2002).
  \bibitem{meson9}
      B.~C.~Hiesmayr,
  Nonlocality and entanglement in a strange system,
  Eur.\ Phys.\ J.\ C {\bf 50}, 73 (2007).
  \bibitem{meson10}
    S.~Banerjee, A.~K.~Alok and R.~MacKenzie,
  Quantum correlations in B and K meson systems,
  arXiv:1409.1034 [hep-ph].
  \bibitem{meson11}
    N.~Nikitin, V.~Sotnikov and K.~Toms,
  Experimental test of the time-dependent Wigner inequalities for neutral pseudoscalar meson systems,
  arXiv:1503.05332 [hep-ph].

  \bibitem{expt1}
  A.~Go [Belle Collaboration],
  Observation of Bell inequality violation in B mesons,
  J.\ Mod.\ Opt.\  {\bf 51}, 991 (2004).
   \bibitem{expt2}
  A.~Go {\it et al.}  [Belle Collaboration],
  Measurement of EPR-type flavour entanglement in Upsilon(4S) ---> B0 anti-B0 decays,
  Phys.\ Rev.\ Lett.\  {\bf 99}, 131802 (2007).
   \bibitem{expt3}
  G.~Amelino-Camelia, 
  {\it et al.},
  Physics with the KLOE-2 experiment at the upgraded DA$\phi$NE,
  Eur.\ Phys.\ J.\ C {\bf 68}, 619 (2010).

\bibitem{neutri1}
M. Blasone, F. Dell'Anno, S. De Siena, and F. Illuminati,  Entanglement in neutrino oscillations, Eur. Phys. Lett. {\bf 85},  50002 (2009).

\bibitem{neutri2}
M. Blasone, F. Dell'Anno, S. De Siena, and F. Illuminati,  Multipartite entangled states in particle mixing, Phys. Rev. D {\bf 77}, 096002 (2008).

\bibitem{neutri3}
   M. Blasone, F. Dell'Anno, S. De Siena, and F. Illuminati,
  On entanglement in neutrino mixing and oscillations,
  J.\ Phys.\ Conf.\ Ser.\  {\bf 237}, 012007 (2010).
  \bibitem{neutri4}
  M.~Blasone, F.~Dell'Anno, S.~De Siena and F.~Illuminati,
  A field-theoretical approach to entanglement in neutrino mixing and oscillations,
  Europhys.\ Lett.\  {\bf 106}, 30002 (2014).
  \bibitem{neutri5}
    M.~Blasone, F.~Dell'Anno, S.~D.~Siena and F.~Illuminati,
  Entanglement in a QFT Model of Neutrino Oscillations,
  Adv.\ High Energy Phys.\  {\bf 2014}, 359168 (2014).

\bibitem{q2flavn}
  A.~K.~Alok, S.~Banerjee and S.~U.~Sankar,
  Quantum correlations in two-flavour neutrino oscillations,
  arXiv:1411.5536 [quant-ph].

\bibitem{Bahcall:2004ut}
  J.~N.~Bahcall, M.~C.~Gonzalez-Garcia and C.~Pena-Garay,
  Solar neutrinos before and after neutrino 2004,
  JHEP {\bf 0408}, 016 (2004).

  \bibitem{Araki:2004mb}
  T.~Araki {\it et al.}  [KamLAND Collaboration],
  Measurement of neutrino oscillation with KamLAND: Evidence of spectral distortion,
  Phys.\ Rev.\ Lett.\  {\bf 94}, 081801 (2005).

  \bibitem{Ashie:2004mr}
  Y.~Ashie {\it et al.}  [Super-Kamiokande Collaboration],
  Evidence for an oscillatory signature in atmospheric neutrino oscillation,
  Phys.\ Rev.\ Lett.\  {\bf 93}, 101801 (2004).

  \bibitem{Michael:2006rx}
  D.~G.~Michael {\it et al.}  [MINOS Collaboration],
  Observation of muon neutrino disappearance with the MINOS detectors and the NuMI neutrino beam,
  Phys.\ Rev.\ Lett.\  {\bf 97}, 191801 (2006).

\bibitem{Abe:2013hdq}
  K.~Abe {\it et al.}  [T2K Collaboration],
  Observation of Electron Neutrino Appearance in a Muon Neutrino Beam,
  Phys.\ Rev.\ Lett.\  {\bf 112}, 061802 (2014).

  \bibitem{Abe:2013fuq}
  K.~Abe {\it et al.}  [T2K Collaboration],
  Measurement of Neutrino Oscillation Parameters from Muon Neutrino Disappearance with an Off-axis Beam,
  Phys.\ Rev.\ Lett.\  {\bf 111}, 211803 (2013).

\bibitem{HMGH}
M.~ Huber, F.~ Mintert, A.~ Gabriel and B.~ C.~ Hiesmayr, Detection of High-Dimensional Genuine Multipartite Entanglement of Mixed States,
Phys. Rev. Lett. {\bf 104}, 210501 (2010).

\bibitem{M90}
N. D. Mermin, Extreme quantum entanglement in a superposition of macroscopically distinct states, Phys. Rev. Lett. {\bf 65}, 1838 (1990).

\bibitem{Svet87}
G. Svetlichny, Distinguishing three-body from two-body nonseparability by a Bell-type inequality, Phys. Rev. D {\bf 35}, 3066 (1987).

\bibitem{chakrabarty}
I. Chakrabarty, P. Agrawal and A. K. Pati, Quantum Dissension: Generalizing Quantum Discord for Three-Qubit States, Eur. Phys. J. D {\bf 65}, 605 (2011).

\bibitem{nfit}
M.~C.~Gonzalez-Garcia, M.~Maltoni and T.~Schwetz,
Updated fit to three neutrino mixing: status of leptonic CP violation,
JHEP {\bf 1411}, 052 (2014).

\bibitem{BTN}
W. Thirring, R.A. Bertlmann, Ph. K\"ohler and H. Narnhofer, Entanglement or separability: The choice of how to factorize the algebra of a density matrix,
Eur. Phys. J. D \textbf{64}, 181 (2011).

\bibitem{HBB}
M. Hillery, V. Buzek, and A. Berthiaume, Quantum Secret Sharing, Phys. Rev. A
\textbf{59}, 1829 (1999).

\bibitem{SHH}
St. Schauer, M. Huber and B. C. Hiesmayr, Experimentally Feasible Security Check for n-qubit Quantum Secret Sharing,
J. Phys. A: Math. Theor. {\bf 43}, 385306 (2010).



\bibitem{lorentzmeasure}
M. Huber, N. Friis, A. Gabriel, Ch. Spengler, and B. C. Hiesmayr, Lorentz invariance of entanglement classes in multipartite systems, EPL {\bf 95}, Number 2, 20002 (2011).





\bibitem{ajoy}
A.~Ajoy and P.~Rungta, Svetlichny's inequality and genuine tripartite nonlocality in three-qubit pure states, Phys.\ Rev.\ A {\bf 81}, 052334  (2010).




\bibitem{CT91}
T.~ M.~ Cover and J.~ A.~ Thomas, {\it Elements of Information Theory} (Wiley-Interscience, 2006).

\bibitem{olivierzurek} H. Olivier and W. H. Zurek, Phys. Rev. Lett. {\bf 88}, 017901 (2001).

\bibitem{henderson}
L. Henderson and V. Vedral,  J. Phys. A {\bf 34}, 6899 (2001).



\bibitem{Hiesmayr:2012}
 B.~C.~Hiesmayr, A.~Di Domenico, C.~Curceanu, A.~Gabriel, M.~Huber, J.~A.~Larsson and P.~Moskal,
 Revealing Bell's Nonlocality for Unstable Systems in High Energy Physics,
 Eur.\ Phys.\ J.\ C {\bf 72}, 1856 (2012).

\bibitem{BH_Hyperon}
B. C. Hiesmayr, Limits Of Quantum Information in Weak Interaction, Sci. Rep. {\bf 5}, 11591 (2015).


\end{thebibliography}


\end{document}